\documentclass[12pt]{article}
\usepackage{graphicx}

\usepackage{amssymb}

\begin{document}



\begin{titlepage}

\title{Gluon Interactions with Color-Neutral Fields}


\author{E.Gilboa\\
Center for Particle Physics, Physics Department,\\
 University of
Texas at Austin, Austin TX 78712-1081 USA}

\maketitle

\begin{abstract}
We investigate the interactions of color neutral fields with
gluons in the regge region, and propose a model in which the field
strength of the gluons couples to these fields.  This model yields
to first order in perturbation theory a structure function which
coincides with that obtained in deep inelastic scattering (in the
Double Log Approximation of QCD) to first order at low $q$ with a
correction. We propose that higher order corrections in this model
will contribute to the parton structure functions beyond the DLA.
It is also shown that the Born approximation in this model yields
a potential having a monopole, and a quadrupole term that may
couple to hadronic currents providing angular momentum transitions
of $\triangle J=2,0$, as is the case in hadron regge trajectories.
\end{abstract}
\end{titlepage}

\section{Introduction}
\label{intro} \label{sec0}

The regge region of QCD poses a great challenge for particle
physicists to this day.  Although great strides in the hard a and
semi-hard regions~\cite{Gribov} have been made, the former sector
is still obscure.  The source of the challenge lies in the
complexity of QCD in the IR region, namely the non- perturbative
sector of QCD when momentum exchanges $q$ are that of the order of
$\Lambda_{QCD}$. One immediate consequence of this complexity is
the inability to describe fundamental interactions among hadrons
using perturbation theory.  To mention a few are hadron
resonances, hadron diffraction dissociation, and hadron structure
(which should have a very rich description when the coupling
between constituent quarks and gluons is very strong). It is also
worth while mentioning that many of these processes are dominated
by an exchange of a color singlet, or what is known as the Pomeron
which has become an entire sub-field of QCD.  This problem has
received great attention~\cite{Lipatov1,Gribov,Levin,Ross}, and
remains elusive because of its non-perturbative description.

Probably one of the most notable achievements of QCD is asymptotic
freedom \cite{Coleman}.  This of course enables perturbation to be
extremely effective in the high energy regime, and is especially
effective in treating the scattering object (which is always a
color neutral field) as made up of an ensemble of very lightly
interacting partons.  Thus one can neglect the interactions among
the various partons within a particular field, and only worry
about the interactions those partons have with fields on which
they scatter from.  Of course as one drops in energy one loses
this nice picture, and compositeness has to be taken into account.
However, here too asymptotic freedom (or the inverse of it) can be
utilized.  Since at lower energies the partons interact very
strongly it is more sensible to treat this ensemble of constituent
fields as one effective color-neutral field (CNF) while enabling
it to interact with constituent fields that do carry color (such
as gluons). In this description a composite color-neutral field
can be viewed as an effective field~\cite{Gribov} composed of
partons whose life time is much greater than their interaction
time. This says that effectively one is in the region where
exchanged momenta between partons is much greater than the mass of
the entire neutral field, or simply put:
\begin{eqnarray*}
\frac{1}{q_{parton}}\approx\triangle t_{int' of parton} \ll
\triangle t _{life}\approx\frac{1}{m_{CNF}}.
\end{eqnarray*}

These scales define to what extent one can treat a composite field
as an effective quantum field without worrying about the vary
difficult problems of specifying interactions between the partons
themselves, let alone their individual interactions with any
external (gluon) fields.  It is evident that such a description
would be particularly useful in a case where CNFs interact at low
exchanged momenta (in the IR region) where the scale of the
interaction may be of the same order as of the masses of the CNFs.
One can now avoid the description among the various partons (since
these will occur at a different higher scale), and still treat the
CNF as a composite field.   If the CNF were to interact with any
color-carrying field, then to first order the CNF may be treated
as neutral parton which must remain neutral at least for a time by
which the interactions take place, namely the life time of the
color-neutral field $\bigtriangleup t_{life}$. In essence this
scheme enables treating interactions of non-local composite fields
in the IR region without specifying the local interactions that
take place between constituent fields such quarks and gluons.
However it is important to remember that once the scale of the
interactions between the various partons is equal to the scale of
the neutral quantum field, such a description is not useful, and a
new scale must be chosen. One therefore expects that any physical
observable obtained from such a description is valid only within
the confines of the scales chosen.
\section{CNF-Gauge Boson Vertex}
\label{sec1}

We propose a model in which a color-neutral field interacts with
two color carrying gauge bosons.  Two gauge particles are needed
as a minimum to preserve the color-neutrality of the CNF.  Of
course this isn't the only interaction possible that could fulfill
this requirement since one could also have interactions of two
CNFs' with multiple gauge fields.  However as we shall argue such
interactions are suppressed significantly.

With this said, amplitudes between {\it{in}} and {\it{out}} states
of two CNFs' and two gauge bosons (see fig.(1)) must fulfill the
following two requirements: the first is due to spin statistics of
the gauge bosons for which the amplitude must satisfy the
following permutation:
\begin{eqnarray*}
A^{\mu\nu}_{ab}(k_1,k_2)=A^{\nu\mu}_{ba}(k_2,k_1).
\end{eqnarray*}

\begin{figure}
\includegraphics[scale=3,width=7cm,height=7cm]{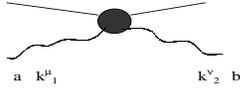}
 \caption{The CNF-gauge boson
vertex} \label{fign0}
\end{figure}
The second is due to current conservation which dictates that the
amplitude should have the Ward identity,
\begin{eqnarray*}
k_{1\mu}A^{\mu\nu}_{ab}=k_{2\nu}A^{\mu\nu}_{ab}=0.
\end{eqnarray*}

These two conditions are suffice to constrain the general form of
such an amplitude to the following:

\begin{eqnarray}
A^{\mu\nu}_{ab}&\sim& \delta_{ab}\Bigg[F_0\left(g^{\mu\nu}k_1\cdot
k_2-k_1^{\nu}k_2^{\mu}\right)\nonumber\\
&+& F_1 \left(q^{\mu}-k_1^{\mu}\left(\frac{q \cdot
k_1}{k_1^2}\right)\right)\left(q^{\nu}-k_2^{\nu}\left(\frac{q\cdot
k_2}{k_2^2}\right)\right)\Bigg] \label{main}.
\end{eqnarray}
with $q^{\mu}=k_1^{\mu}-k_2^{\mu}$.
\newcommand{\M}{(g^{\mu\nu}k_1\cdot
k_2-k_1^{\nu}k_2^{\mu})}

Since the amplitude above really represents a color neutral
current then it follows that $F_0$, and $F_1$ are associated with
the structure functions of the CNF. The amplitude~(\ref{main})
says that $F_1$ should vanish as $q$ gets small.  In fact the term
proportional to $F_1$ becomes relevant at high $q$.  In such a
scenario, our assumption of having the momentum of the interaction
of the CNF with any external field not being comparable to the
momentum of the interactions between the partons (which make up
the CNF) may be compromised. As was indicated, when these two
scales are of the same order our description of a CNF becomes
invalid. In light of this we shall focus only the term
proportional to $F_0$ in what is to follow.

In searching for a field theoretical model which would give rise
to such a structure of the scattering amplitude, a particular
gauge invariant term added to the total Lagrangian of the
non-abelian gauge theory fulfills the above requirements, and is
given by
\begin{eqnarray}
{\mathcal{L}}_{CNF-g}=\bar{\psi}\Bigg(\frac{\lambda}{2}{\mathcal{F}}^{\mu\nu}_{ab}{\mathcal{F}}_{\nu\mu
ab}+...+\frac{\lambda_n}{2+2n}\Big({\mathcal{F}}^{\mu\nu}_{ab}{\mathcal{F}}_{\nu\mu
ab}\Big)^n\Bigg)\psi. \label{Lagran}
\end{eqnarray}
Where ${\mathcal{F}}^{\mu\nu}_{ab}$ is the field strength of the
gauge field and $\psi$ is a color neutral field.

Of course apart from one of the first term in~(\ref{Lagran}) this
interaction does not directly produce the kind of interaction
given in~(\ref{main}), but yields for the $n^{\rm{th}}$ term
$3+2n$ vertices each having two CNFs and $\gamma(1+n)$
($\gamma=2,3,4$) gauge fields. Because~(\ref{Lagran}) is
manifestly gauge invariant, for each vertex in each term
of~(\ref{Lagran}) one can integrate out $\gamma(1+n)-2$ gauge
fields to produce~(\ref{main}).  So really it is only the first
term in~(\ref{Lagran}) that is relevant since the other terms are
proportional to it up to a coupling constant, which could as well
be redefined. We shall not present a formal proof to this claim,
but instead provide an argument based on dimensional analysis for
why we can neglect higher order terms in~(\ref{Lagran}).

The coupling $\lambda_n$ has dimensions of
$(mass_{\rm{CNF}})^{-(d+4n)}$, where $d$ depends on the space-time
properties of the field (for example $d=3$ for a spinor field
while for a scalar field $d=2$), while the mass has a natural
cut-off scale at $\Lambda_{QCD}$. The negative mass dimension of
the coupling indicates that at high momentum
transfer~(\ref{Lagran}) is comprised of what are so called
irrelevant operators~\cite{Peskin}.  It can be seen
from~(\ref{Lagran}) that the $n^{\rm{th}}$ amplitude is
proportional (in the $t$-channel) to
$\sim\left(\frac{\sqrt{t}}{\Lambda_{QCD}}\right)^{d+4n}$, which
means they drop as $t$ goes down.  As a consequence it is apparent
that
\begin{displaymath}
(A_{ab}^{\mu\nu})_{n+1}=\left(\frac{\sqrt{t}}{\Lambda_{QCD}}\right)^4
(A_{ab}^{\mu\nu})_n.
\end{displaymath}
If the term in parenthesis above is small to begin with then the
$n^{\rm{th}}+1$ amplitude will be significantly dwarfed from the
previous one.  Hence for small $t$ it is suffice to consider the
first term in the series of~(\ref{Lagran}).  This argument also
establishes that the more gluons appear in vertices with CNFs, the
faster those amplitudes associated with these vertices will
vanish. Therefore the vertex with the minimum amount of gluons
(which is two) would give the largest contribution to the total
amplitude.  The first term in~(\ref{Lagran}) is also of particular
interest (as will be shown) when the CNF can be described by a
`manyfield' \cite{Neeman}, one which could describe a collection
of states that have different spin (like a regge trajectory). This
term may facilitate transitions between states that differ in
angular momentum $\triangle J=2,0$ in the Born approximation.

\section{QCD Corrections to CNF-Gluon Vertex}

Given the first term in~(\ref{Lagran}) we seek to evaluate the
amplitudes of the following process:
\begin{displaymath}
 \rm{CNF}+ g\rightarrow \rm{CNF} + g.
\end{displaymath}

\begin{figure}
\includegraphics[scale=1.2,width=10cm,height=14cm,totalheight=10cm]{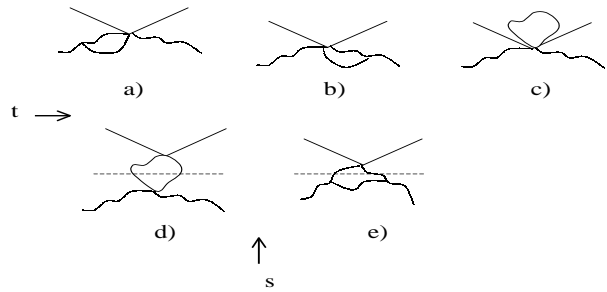}
\caption{First order corrections to the CNF-gluon vertex}
\label{fign1a}
\end{figure}

We wish to evaluate corrections to this process that are first
order in $\lambda$, and in $\alpha_s$ \footnote{It is important to
stress that interactions in this model which include pure QCD
exchanges up to first order in $\lambda$ have no bearing on the
running of $\alpha_s$ which is controlled by ${\mathcal{L}}_{QCD}$
proper.}. There are five relevant terms in the perturbation series
contributing to the correction of the vertex given by the first
term of~(\ref{Lagran}), and are described by the Feynman diagrams
shown in figures~(\ref{fign1a}a-\ref{fign1a}e). These
contributions can be grouped into two categories: the first
(figs.~(\ref{fign1a}a-\ref{fign1a}c)) are terms that result in the
reduction of the three gluon vertex, and the four gluon vertex to
a two gluon vertex respectively.  These will be shown to have no
physical relevance since they exactly correspond to integrating
out the extra gluon fields as discussed in section (\ref{sec1}).
The second group (figs.~(\ref{fign1a}d, \ref{fign1a}e)) do have a
physical contribution, and are `genuine' QCD corrections which
contribute to $F_0$.

We compute these amplitudes by scaling the gauge field by the
familiar prescription $A^{\mu}\rightarrow gA^{\mu}$, and
anticipate the strong coupling to show in propagators.  For these
corrections it is also assumed that the CNFs are scalars, however
this can be generalized to a CNF of any spin without affecting the
results that follow from our calculations.

Since we are interested in terms proportional to $F_0$ it is worth
while to put the external gluons on shell since this will
eliminate terms proportional to $k_1^{\mu}, k_2^{\nu}$, and leave
only terms proportional to $k_1^{\nu}, k_2^{\mu}$ as
in~(\ref{main}).  For the first group of diagrams (in the Feynman
gauge figs.~(\ref{fign1a}a-\ref{fign1a}c)) this will introduce a
`temporary' infra-red divergence for which the off-shell gluons
are assigned a fictitious mass.

The first of these is given by:
\begin{eqnarray}
ic_{1~ab}^{\mu\nu} =
-\frac{\lambda}{4}4\pi\int\frac{d^dk}{(2\pi)^d}\alpha_s\Gamma^{\mu\sigma\rho}_{acd}(k_1,k)
\frac{g^{\sigma\sigma'}\delta_{cc'}}{k^2-\mu^2}\times\nonumber\\
\frac{g^{\rho\rho'}\delta_{dd'}}{(k-k_1)^2-\mu^2}
\Gamma^{\rho'\sigma'\nu}_{c'd'b}(k_1,k_2,k)\label{c1},
\end{eqnarray}
where $\Gamma$ \cite{Peskin} is the familiar $SU(N)$ triple gauge
boson vertex.

Since diagrams contributing the running of $\alpha_s$ have been
omitted (after all these are also first order in $\alpha_s$) a
legitimate questions is raised on whether the expression for the
running of $\alpha_s$ should be included in~(\ref{c1}), and thus
integrated out.  The answer to this question depends on the
structure of the diagram. As can be seen from
figure~(\ref{fign1a}a), the off-shell propagators are directly
connected to the vertex~(\ref{main}). Thus, varying the strong
coupling implies a variation in $\lambda$, or simply put
$\lambda=\lambda(\alpha_s(k))$.  This in itself is a true
statement (this is evident from~(\ref{Lagran}) when the gauge
field is unscaled by $g$), though a variation of $k$ implies a
variation in $\lambda$ which would be inappropriate at this stage
since we are performing perturbation to first order in $\lambda$
where it is assumed to be constant; hence $\alpha_s$ is constant
as well. This argument applies to all but the fifth diagram.

Having cleared this issue~(\ref{c1}) is given by:
\begin{eqnarray}
ic_{1~ab}^{\mu\nu}& = & -\delta_{ab}\frac{N\lambda
\alpha_s}{4}4\pi\int^1_0dx\int\frac{dl^d}{(2\pi)^d}\frac{-3\M+2g^{\mu\nu}(\frac{1-d}{d})l^2}{(l^2-\mu^2)^2}\nonumber\\
& = & i\delta_{ab}\frac{\pi N\lambda
\alpha_s}{(4\pi)^{\frac{d}{2}}}g^{\mu\nu}(1-d)\nonumber\left(\frac{1}{\mu^2}\right)^{1-\frac{d}{2}}\\
&+&i\delta_{ab}\frac{3\pi
N\alpha_s}{4(4\pi)^{\frac{d}{2}}}\M\Gamma\left(2-\frac{d}{2}\right)
\left(\frac{1}{\mu^2}\right)^{2-\frac{d}{2}}. \label{c1a}
\end{eqnarray}

The second amplitude~(fig.~(\ref{fign1a}b)) $c_{2~ab}^{\mu\nu}$ is
obtained from $c_{1~ab}^{\mu\nu}$ on interchanging $k_1\rightarrow
k_2$, $\mu\rightarrow\nu$, and $a\rightarrow b$. The latter is
symmetric upon these interchanges hence
$c_{2~ab}^{\mu\nu}=c_{1~ab}^{\mu\nu}$.

Evaluating the third amplitude (fig.~\ref{fign1a}c) we get:
\begin{eqnarray}
ic_{3~ab}^{\mu\nu}& =& -\frac{N\alpha_s\delta_{ab}}{2}\int\frac{d^dk}{(2\pi)^d}\frac{(d-1)g^{\mu\nu}}{k^2-\mu^2}\nonumber\\
&= &i \frac{2 \pi N
\alpha_s\delta_{ab}}{(4\pi)^{\frac{d}{2}}}g^{\mu\nu}(d-1)\left(\frac{1}{\mu^2}\right)^{1-\frac{d}{2}}
\end{eqnarray}.

Summation on the first three terms gives:
\begin{eqnarray}
\sum^{3}_{i}
c_{i~ab}^{\mu\nu}=\frac{3N\alpha_s}{8\pi}\bigtriangleup_{ab}^{\mu\nu}(k_1,k_2)\left(\frac{2}{\epsilon}-\log{\mu^2}-\gamma
+\log{4\pi}\right)\label{sum1}
\end{eqnarray}
where $\bigtriangleup_{ab}^{\mu\nu}(k_1,k_2)=\delta_{ab}\M$.

This sum suffers from the familiar ultraviolet divergences,
however it does not depend on any dynamical variables and
therefore does not contribute to $F_0$, at least not to first
order in $\alpha_s $.  One can subtract these divergences by
redefining $\lambda$ with counter terms, and impose the following
condition:
\begin{eqnarray}
F_0(t=t_o,\mu\rightarrow 0)=\lambda\label{con1},
\end{eqnarray}
where $F_0$ is given by:
\begin{eqnarray}
F_0(t,\mu)=\frac{\epsilon_{a\mu}\epsilon_{b\nu}}{G(k_1,k_2)}\left(\lambda\bigtriangleup^{\mu\nu}(k_1,k_2)
+\sum^5_i  c_{i~ab}^{\mu\nu}(t,\mu)\right) + \delta F_0
\label{sum2},
\end{eqnarray}
with
$G(k_1,k_2)=\epsilon^a_{\mu}\epsilon^b_{\nu}\bigtriangleup_{ab}^{\mu\nu}(k_1,k_2)$\footnote{Indices
pertaining to polarization are suppressed.}, and $t_o$ being a
renormalization scale chosen at some momentum exchange.

The function $F_0$ is a CNF structure function obtained from pure
QCD processes, and includes contributions from all five diagrams,
while the term $\delta F_0$ contains the appropriate counter
terms.

Proceeding to evaluate the second group of diagrams we note that
unlike the first three the former~
(fig~(\ref{fign1a}d,\ref{fign1a}e)) are most easily obtained
(especially in the regge region) using $s$-channel unitarity,
namely the Cutkosky cutting rules~\cite{Cutkosky}. Once the
$s$-channel amplitude is obtained, one then can use crossing
symmetry to get the $t$-channel amplitude which presently is of
most interest.

According to the cutting rules each diagram~
(figs.~(\ref{fign1a}d, \ref{fign1a}e)) is split into two tree
level amplitudes; one giving the process of $g+g\rightarrow g+g$,
the other giving the process of $g+g\rightarrow \rm{CNF+CNF}$.

One can now evaluate the imaginary part for the fourth and fifth
amplitudes using these cuts to give:
\begin{eqnarray}
{\mathcal{I}}m(c_{i~ab}^{\mu\nu})_{s-channel} & = &
\frac{g^4}{2}\sum_{g}\int\frac{d^4 u}{(2\pi)^3}\frac{d^4
u'}{(2\pi)^3}\delta(u^2)\delta(u'^2)(2\pi)^4\delta^4(k_1+k_2-u_i-u_i')\nonumber\\
 & \times &
 (Y_{abcd}^{\mu\nu\sigma\rho})_i(\epsilon_{\sigma c}\epsilon_{\sigma' c'}
 \epsilon_{\rho~d}\epsilon_{\rho'~d'})
 (X_{c'd'}^{\rho'\sigma'})_i^{\dagger} \label{c45},
\end{eqnarray}
where $u_i, u'_i$ are the momenta of the out-going gluons for each
diagram respectively, and the sum is over gluon
polarizations\footnote{Because the gauge fields have been scaled
by $g$ then polarizations should be scaled as well meaning,
$\epsilon^{\mu}\rightarrow g\epsilon^{\mu}$ which explains the
factor of $g^4$ in front of~(\ref{c45}).}.

The terms $Y_i,X_i$ are the scattering amplitudes of the $g,g$,
and $g,CNF$ processes respectively.  For the fourth amplitude
these tree level expressions are given by:
\begin{eqnarray}
(Y^{\rho\sigma\mu\nu}_{abcd})_4&=&\frac{-1}{g^2}(\Gamma_4)^{\rho\sigma\mu\nu}_{abcd}(k_1,k_2,u'_4,u_4)\\
(X^{\rho'\sigma'}_{c'd'})_4&=&\lambda\bigtriangleup^{\rho'\sigma'}_{c'd'}(u'_4,u_4)
\label{c4a} ,
\end{eqnarray}
where $\Gamma_4$ is the familiar four gluon vertex, and
$u_4=\frac{k_1+k_2}{2}-k$, $u'_4=\frac{k_1+k_2}{2}+k$.

While for the fifth amplitude the tree level expressions are:
{\begin{eqnarray} (Y_{abcd}^{\mu\nu\sigma\rho})_5 & = &
4\pi\alpha_s
\Gamma^{\mu\sigma\lambda}_{aec}\Gamma^{\hspace{0.05cm}\nu\rho}_{\lambda\hspace{0.05cm}bed}
\left(\frac{1}{k^2}\right)\\
(X_{c'd'}^{\rho'\sigma'})_5 &
=&\lambda\bigtriangleup^{\rho'\sigma'}_{d'c'}(u'_5,u_5)\label{c5a},
\end{eqnarray}
where $u_5 =k_1-k$, $u'_5=k_2+k$.

Working in the center of mass frame with the in-coming gluons
momenta given by:
\begin{displaymath}
k_1=(\frac{\sqrt{s}}{2},\frac{\sqrt{s}}{2},0,0)\hspace{1cm}
k_2=(\frac{\sqrt{s}}{2},-\frac{\sqrt{s}}{2},0,0),
\end{displaymath}
it is convenient to parameterize the vector $k$ with Sudakov
variables:
\begin{equation}
k^{\mu}=\alpha k^{\mu}_1+\beta k^{\mu}_2+ {\mathbf{k}}.
\label{Sud}
\end{equation}

This parametization is particularly convenient where soft
interactions (regge region) in the $t$-channel take place since
they make $k$ predominantly transverse, and that implies
$\alpha,\beta\ll1$.  In what follows this enables second order
terms in these variables to be neglected.

With this methodology one notices that in the case of
$(c_{4ab}^{\mu\nu})_{s-channel}$ neither amplitudes of the cut
diagram~(fig.~(\ref{fign1a}d)) contains any off-shell gluon
propagators, and all gluons are on shell. Hence this amplitude is
most easily obtained using a physical gauge \cite{Gribov2} where a
gluon has two independent degrees of freedom with two polarization
vectors that obey the following relation:
\begin{equation}
\sum_G\epsilon^{\mu}\epsilon^{\nu}=d_{\perp}^{\mu\nu},
\end{equation}
and the sum is over gluon polarizations.

The tensor $d^{\mu\nu}_{\perp}$ is the transverse metric
$(0,0,-1,-1)$.  This tensor is obtained when the polarization
vectors of the gauge fields in~(\ref{c45}) are chosen to be purely
along the transverse plane with respect to the in-coming gluon
momenta.

Utilizing this gauge freedom, implementing (\ref{c45}), and
keeping only first order terms in Sudakov parameters the imaginary
amplitude of $(c_{4ab}^{\mu\nu})_{s-channel}$ becomes:
\begin{eqnarray}
{\mathcal{I}}m(c_{4~ab}^{\mu\nu})_{s-channel}&=&-\frac{s}{4\pi}\int
d\alpha d\beta{\mathbf{d^2k}}\delta(u^2)
\delta(u'^2)\nonumber\\
& \times &
N\lambda\alpha_s(\Gamma_4)^{\rho\sigma\mu\nu}_{abcd}d_{\perp}^{\rho\rho'}d_{\perp}^{\sigma\sigma'}
\delta_{cc'}\delta_{dd'}\bigtriangleup^{\rho'\sigma'}_{d'c'}\nonumber\\
&&= -\frac{N\lambda}{2\pi s
}\bigtriangleup^{\mu\nu}_{ab}(k_1,k_2)\int^{\sqrt{s}}_0{\mathbf{d^2k}}\alpha_s,\label{c4b}
\end{eqnarray}
The upper limit on the integral in~(\ref{c4b}) is required since
the out-going gluons in the cut amplitude are on shell.

It is important to reiterate that in~(\ref{c4b}) too $\alpha_s$ is
fixed. The two propagators flow into the (fixed) $\lambda$
dependent $g-\rm{CNF}$ vertex, which constrains $\alpha_s$ not to
run. Thus amplitude~(\ref{c4b}) is finite, and is given by:
\begin{eqnarray}
{\mathcal{I}}m(c_{4~ab}^{\mu\nu})_{s-channel}=-\M\frac{N\lambda\alpha_s}{2}\label{c4c}
\end{eqnarray}
Using the analytical properties of the $S$-matrix \cite{Collins}
the real part of this amplitude is given by:
\begin{eqnarray}
{\mathcal{R}}e(c_{4~ab}^{\mu\nu})_{s-channel}=\M\frac{N\lambda\alpha_s}{2\pi}\ln\left(\frac{s}{|t|}\right)\label{c4d}.
\end{eqnarray}
The $t$-channel amplitude is purely real (for $t<0$), and is
simply obtained by $s,t$ crossing symmetry.

The imaginary amplitude~(fig.~(\ref{fign1a}e)) is obtained in a
similar fashion though here we choose to work in the Feynman gauge
for the off-shell gluon propagator.  Applying the cutting rules,
and again neglecting second order terms in Sudakov parameters it
follows that:
\begin{eqnarray}
{\mathcal{I}}m(c_{5~ab}^{\mu\nu})_{s-channel}= \M2\lambda N
\int^{\sqrt{s}}_{q_o}
\frac{d^2{\mathbf{k}}}{4\pi}\frac{\alpha_s}{{\mathbf{k}}^2
}.\label{c5b}
\end{eqnarray}

Unlike the first four amplitudes already evaluated, the
amplitude~(\ref{c5b}) is distinct in two ways.  First, we have
chosen to use a cut-off for the integral as a lower bound instead
of introducing a mass parameter. This is because that unlike the
first three amplitudes, the infra-red divergence appearing
in~(\ref{c5b}) is not `temporary', but is a result of the
non-perturbative nature of QCD at low energies. Even though we
have assumed the gluons to be on shell, one could justify this
cut-off \cite{Ross} by noting that in a more realistic scenario
gluons never appear as asymptotically free particle states (due to
their confinement), and therefore will posses some off-shell
momenta.  Second, the coupling $\alpha_s(k)$ can be set to run.
The coupling that appears inside the integral of~(\ref{c5b}) is
due to the gluon propagator that appears in the $t$-channel, which
is completely detached from the $g$-CNF vertex.  Thus one can keep
$\lambda$ fixed because the momentum flowing into the vertex,
namely $k_1+k_2$ does not depend on $k$.

Integrating (\ref{c5b}), and again using the analytical properties
of the amplitude, the real part of $c^{\mu\nu}_{5~ab}$ is given
by:
\begin{equation}
{\mathcal{R}}e(c_{5~ab}^{\mu\nu})_{s-channel}=-\M
\frac{1}{2}f(s,q_o) \ln{\left(\frac{s}{|t|}\right)},\label{c5c}
\end{equation}
where
\begin{displaymath}
f(s,q_o)=\frac{8\lambda N}{\pi b
}\left(\ln{\ln{\left(\frac{s}{\Lambda^2_{QCD}}\right)}}-
\ln{\ln{\left(\frac{t_o}{\Lambda^2_{QCD}}\right)}}\right).
\end{displaymath}
and $b=\frac{11}{3}N-\frac{2}{3}n_f$.\\

Similarly the $t$-channel amplitude is obtained by $s,t$ crossing symmetry.\\

Using condition~(\ref{con1}) together with~(\ref{sum2}) in the
$t$-channel, the following is obtained:
\begin{eqnarray*}
\delta
F_0&=&-\frac{\epsilon_{a\mu}\epsilon_{b\nu}}{G(k_1,k_2)}\sum^5_i
 c_{i~ab}^{\mu\nu}(t=t_o,\mu\rightarrow0)\\
&=&-\frac{\epsilon_{a\mu}\epsilon_{b\nu}}{G(k_1,k_2)}\sum^3_i
 c_{i~ab}^{\mu\nu}-\frac{N
 \lambda\alpha_s}{2\pi}\ln{\left(\frac{|t_0|}{s}\right)},
\end{eqnarray*}

and with the tree level vertex it follows that:
\begin{eqnarray}
F_0=\lambda\left[1+\frac{1}{2}f(t,t_o)\ln{\left(\frac{s}{|t|}\right)}
 +
 \frac{N\alpha_s}{2\pi}\ln{\left(\frac{t}{t_o}\right)}\right].\label{cver}
\end{eqnarray}

The first three amplitudes in the series get completely cancelled
by the $\delta F_0$, and what are left are the last two terms that
contribute to the CNF's structure function.  This is an
affirmation to our claim maid in section~(\ref{sec1}), namely
amplitudes associated with integrating out extra gluon fields from
vertices that contain multiple gluon interaction with CNFs have no
physical contribution.  This also establishes that terms
proportional to $\bigtriangleup_{ab}^{\mu\nu}(k_1,k_2)$ are
relevant at low energy since the terms proportional to $F_1$
in~(\ref{main}) have been suppressed by second order Sudakov
parameters in the last two diagrams.  These results coincide with
our discussion in section~(\ref{sec0}) regarding the momentum
scale at which the CNFs interact.  To clarify this, it is
important to note that the tensor
$\bigtriangleup_{ab}^{\mu\nu}(k_1,k_2)$ arises as a part of the
kinetic term of the gluons in~(\ref{Lagran}) which is proportional
to $(\partial^{\mu}A_a^{\nu}-\partial^{\nu}A_a^{\mu})^2$.   This
term is a measure of the variation of the gauge field in space and
time, and thus describes a non-local quantity.  At low $q$ it is
very well known that quantum fluctuations become quite strong (as
$\alpha_s$ rises).  Thus, if $q$ is of the same order as the mass
of the CNF, or in other words when quantum fluctuations are of the
same order as the size of the CNF, then variation of the gauge
field in space and time would be significant.  Since at low
momentum interactions are mediated by non-local structures,
explains why terms proportional to
$\bigtriangleup_{ab}^{\mu\nu}(k_1,k_2)$ dominate in this energy
region. On the other hand the tensor proportional to $F_1$ arises
from higher order terms in the perturbation series and describes
interactions of the local constituents fields (quarks, gluons)
given by terms in the conventional QCD Lagrangian. Therefore these
terms dominate at scales which are higher compared to that of the
CNF.

The first two terms in~(\ref{cver}) coincide exactly with the
first two terms of the parton structure function obtained in
semi-hard processes in QCD
\cite{Gribov,Gross,Dokshi,Lipatov2,Lipatov3} in the Double Log
Approximation (DLA) . In this approach QCD structure functions are
obtained purely within the realm of perturbative QCD when summing
on ladder diagrams.  These are given by \cite{Gribov}:
\begin{eqnarray}
F=\sum^{\infty}_{n=0}\frac{(f(t,t_o))^n
\left(\ln{\left(\frac{1}{x}\right)}\right)^n}{2^n
(n!)^2}\label{qsf},
\end{eqnarray}
where $x=\frac{|t|}{s}$.

One should not expect though that our model will replicate this
result since the method by which the latter was obtained was
through cutting diagrams going downward the gluon ladder.  In the
current model cuts can only be made across (going sideways) the
ladder. However since our model coincides with the DLA to first
order at low $q$, indicates that higher order corrections will
have new contributions to parton structure functions as $q$ drops
(the third term in~(\ref{cver}) is one such correction).  We base
this conjecture further on noticing that the ladder cuts in this
model can lead to the reggeization \cite{Lipatov,Gribov} of the
$t$-channel gluon (fig.~(\ref{fign1a}e))\footnote{Work in
preparation.}.  This would make the parton structure function at
low $q$ be dominated by local exchanges of reggeized gluons, which
would lead to its reggeized form, namely $F_0\sim |t|^{\alpha}$,
where $\alpha$ is related to the Pomeron's regge trajectory
intercept \cite{Ross}.

\section{Born Approximation and Angular Momentum Transitions}
\label{sec2}

In the context of our model we look at the behavior of the
scattering amplitude at low $t$ (to zeroth order in $\alpha_s$) as
to extract a potential like behavior for when a color neutral
field interacts with a background gluon field in a colorless
state. The situation may be applicable to hadrons interacting with
heavy nuclei where hadronic densities are large enough to produce
such background fields.  The scenario is such that $R_h\ll r$,
where $R_h$ is the effective radius of the hadron, and $r$ is the
radius of the region where such background fields exist. This can
describe a scattering process of a light hadron off a heavy nuclei
(or nuclear matter), or an interaction where these two may form a
bound state.  In both cases the interaction may be considered soft
enough to leave the hadrons intact.

\begin{figure}
\includegraphics[scale=2,width=12cm,height=10cm,totalheight=10cm]{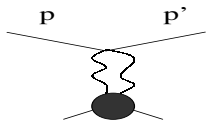}
\caption{Diagram contributing to the Born Approximation}
\label{fign2}
\end{figure}

The interaction in question is shown in figure~(\ref{fign2}).
Since the scattering of the CNF is with a colorless state the
gluons are off-shell.  What is said in effect is that the two
gluons emerging from the upper vertex~(fig.~(\ref{fign2})) may
split independently to form `fan' diagrams \cite{Gribov}, but will
eventually form a colorless state.  Further, since the interaction
is a soft one, it can be assumed that the initial and final energy
of the CNF is approximately equal, and the gluons posses the same
momentum (flowing in opposite directions) given by:
\begin{equation}
k_1=-k_2=\frac{p-p'}{2}.
\end{equation}

With these approximations the amplitude can be given by the
following:
\begin{equation}
<p'|iT|p>=iV_{ab}^{ij}(k)M_{ab ij} \label{amp1},
\end{equation}
where
\begin{equation}
V^{ij}_{ab}=\lambda\delta_{ab}\frac{{\mathbf{k}}^i{\mathbf{k}}^j+\delta^{ij}|{\mathbf{k}}|^2}
{|{\mathbf{k}}|^4},\label{potena}
\end{equation}
and ${\mathbf{k}}=\frac{{\mathbf{p}}'-{\mathbf{p}}}{2}$.

The term $M_{ab ij}$ is the amplitude arising from colorless
states appearing on the bottom of the diagram (fig.~\ref{fign2})
which in our approximation are independent of ${\mathbf{k}}$ and
consist of higher order terms in QCD.  In effect $V^{ij}_{ab}(k)$
is the potential arising from a process where a color neutral
field interacts with a digluon colorless state.

Taking its Fourier transform the following is obtained:
\begin{eqnarray}
V^{ij}_{ab}(x)=-\lambda\delta_{ab}\left(\nabla^2+
\partial^i\partial^j\right)\int
\frac{d^3{\mathbf{k}}}{(2\pi)^3}\frac{e^{i{\mathbf{k}}\cdot{\mathbf{x}}}}{|{\mathbf{k}}|^4}\nonumber\\
=\lambda\delta_{ab}\frac{1}{3\pi
r}\left(\delta^{ij}-\frac{1}{8r^2}Q^{ij}\right),\label{potenb}
\end{eqnarray}
$Q^{ij}=3x^ix^j-\delta^{ij}r^2$.

The potential $V^{ij}_{ab}$ has split into two components; a
monopole term, and a quadrupole term.  This splitting has
significance in the context of excited states of nucleons, or what
are known as Regge trajectories.  It is observed
\cite{Barnes,Chew} that in a $J/m^2$ plot there are transitions of
angular momentum $\triangle J=2,0$ between hadronic states.  In
addition, there are algebraic models \cite{Neeman,Neeman2} which
utilize symmetries of hadronic states requiring volume
conservation ($SL(3R)$, $SL(4R)$ spectrum generating algebras
derived from confinement) that give such selection rules for these
hadronic excitations. Such selection rules are derived from
hadronic currents that are related to a "shear" tensor, and a
"dilaton" scalar given by:
\begin{eqnarray}
T^{\mu\nu}&=&\int d^3x
\hspace{0.1cm}x^{\mu}\theta^{0\nu}+x^{\nu}\theta^{0\mu}-\frac{1}{4}g^{\mu\nu}x^{\sigma}\theta_{0\sigma}\\
D&=&\frac{1}{4}\int d^3x
\hspace{0.1cm}x^{\sigma}\theta_{0\sigma}\label {shd},
\end{eqnarray}
where $\theta^{\nu\mu}$ is the energy momentum tensor for hadronic
fields derived from QCD \cite{Neeman}.

The shear tensor $T^{\mu\nu}$ is a traceless symmetric tensor. It
contains a tensor $T^{ij}$ which transforms as a spherical tensor
of rank two under rotations, and therefore has a non-zero
expectation value between the states
\begin{displaymath}
<J|T^{ij}|J\pm 2>\neq 0.
\end{displaymath}

This tensor provides excitations along a specific regge trajectory
which stems from hadronic structure deformation with volume
conservation.  Thus it should couple to the quadrupole term in
(\ref{potenb}).

The tensor $D$ is a dilaton which transforms as a scalar under
rotations, and therefore has a non-zero expectation value between
states of the same angular momentum namely:
\begin{displaymath}
<J|D|J>\neq 0.
\end{displaymath}

This tensor describes dilatations of hadronic fields, and
therefore provides transitions between different regge
trajectories; no volume conservation.  Therefore it will couple to
the monopole term in~(\ref{potenb}).

It is important to note that this potential can also arise for a
field strength of an abelian gauge theory (electromagnetic field)
coupled to a CNF.  However following the discussion in section
~(\ref{sec1}), amplitudes in this model are proportional to $A\sim
\left(\frac{q}{\Lambda}\right)^d$. This means that in the regge
region a QED process will be suppressed by $\Lambda_{QED}$ which
is much greater than $q$, and therefore will give terms that
vanish.  For QCD, $q$ and $\Lambda_{QCD}$ may be of the same order
(in the IR region) producing a non-vanishing amplitude.

\section{Conclusion}

The use of the field strength in the IR region as a field
theoretical tool to explain phenomena at low $q$ has already been
utilized in other models especially concerning that of the Pomeron
\cite{Levin}.  In this non-perturbative model Kharzeev and Levin
have shown that the trace of the field strength (\ref{Lagran}) is
directly proportional to the trace of the QCD energy momentum
tensor which at small momentum (and assuming chiral symmetry), is
proportional to the pion field and its momentum.  Thus at long
distances the two emerging gluons (by which the CNFs' interact)
hadronize to produce a pion in a first order approximation; a
non-local process which is a direct result of the field strength
interaction.  Further, at low momenta it is believed that QCD
gluon fields may be described by instantons. In obtaining these
semi-classical solutions for the field equations \cite{Ryder}, one
defines a four volume on which $F^{\mu\nu}_{ab}$ is defined.
Boundary conditions are then imposed on the three surface for the
field strength rather than for the gauge field. It follows that in
this scheme the field strength is taken to be a non-local object
which effectively describes the semi-classical fluctuations of the
gauge fields.  All these models coincide with the picture that at
low momentum the parton's wave function becomes more spread and
less localized due to the rise of the strong coupling.  Since our
model deals with color-neutral fields defined to be as non-local
entities as a priori, it is only natural that some form of the
field strength should play a role in describing interactions among
these fields. Although our model is similar to the schemes
mentioned above by the inclusion of the field strength at low
energies, it differs from them in that the treatment here was
perturbative, and thus may provide a bridge between the
perturbative and the non-perturbative sectors of QCD.



\end{document}